\documentstyle[epsfig]{mn}  

\begin{document}  
  
\title[]{Energetic particle acceleration in a 3D magnetic field  
reconnection model: a role of MHD turbulence}

\author[T. Kobak \& M. Ostrowski]{T. Kobak$^*$ \& M. Ostrowski \\  
Obserwatorium Astronomiczne, Uniwersytet Jagiello\'{n}ski, ul. Orla 171,  
30-244 Krak\'{o}w, Poland \\( $^*$E-mail: kto{@}oa.uj.edu.pl)  }  
  
\date{}  
  
\maketitle  
  
\begin{abstract}  
The role of MHD turbulence in the cosmic ray acceleration process in a
volume with a reconnecting magnetic field is studied by means  
of Monte Carlo simulations. We performed modelling of proton  
acceleration with the 3D analytic model of stationary reconnection of  
Craig et al. (1995) providing the unperturbed background conditions.  
Perturbations of particle trajectories due to a turbulent magnetic field  
component were simulated using small-amplitude pitch-angle  
momentum scattering, enabling modelling of both small and large  
amplitude turbulence in a wide wave vector range. Within the approach,  
no second-order Fermi acceleration process is allowed.  
Comparison of the acceleration process in models  
involving particle trajectory perturbations to the unperturbed one  
reveals that the turbulence can substantially increase the acceleration  
efficiency, enabling much higher final  
particle energies and flat particle spectra.  
\end{abstract}  
  

\section{Introduction}  
  
Particle acceleration processes in regions of magnetic field 
reconnection can play an important role in various astrophysical sites 
by providing high energy particles, heating thermal plasma and by 
changing the magnetic field configuration (e.g. Somov 1994). The first 
of these processes, accelerating particles to cosmic ray energies, acts 
due to electric fields occurring in the reconnection region. Two 
approaches were applied to describe the acceleration process. In the more 
basic one, analytic or semi-analytic models for the reconnection region 
provided background for the derivation of particle trajectories and 
discussion of the energy distribution of particles escaping from the 
reconnection volume. A simple discussion of particle trajectories 
presented by Speiser (1965; see also Sonnerup 1971) shows that even a 
tiny magnetic field component near the neutral current sheet 
efficiently removes ergetic particles and decreases its final energy. 
Further studies by Stern (1979) and Wagner et al. (1981) revealed the 
importance of the O-type neutral line regions for the acceleration, due 
to partial particle trapping in the accelerating volume. Deeg et al. 
(1991) and Somov \& Kosugi (1997) discussed several features of the 
acceleration process in the X-type stationary reconnection. The time 
dependent effects of the acceleration process were introduced by 
considering the background magnetic and electric fields as derived from 
simple MHD simulations (cf. Sato et al. 1982, Scholer \& Jamitzky 1987, 
Atkinson et al. 1989, Zelenyi et al.  1990). 
  
Substantial progress in considering realistic reconnection processes 
came from the two dimensional ($\equiv$ 2D; Matthaeus et al.  1984, 
Ambrosiano et al. 1988, Scholer \& Jamitzky 1989, Veltri at al.  1998) 
and 3D (Birn \& Hesse 1994, Schopper et al. 1999, Kliem et al.  1998) 
MHD modelling involving a perturbed magnetic field as the initial 
condition. This approach yields a complex structure with X-type and 
O-type null points and magnetic field perturbations of scales 
comparable to the macroscopic structure dimensions. Ambrosiano et al. 
(1988) show that such a turbulent neutral point mechanism influences the 
acceleration process in two ways. It enhances the reconnection magnetic 
field while producing a stochastic electric field that gives rise to 
momentum diffusion, and it also produces magnetic bubbles and other 
irregularities that can temporarily trap test particles in the strong 
reconnection field for times comparable to the magnetofluid 
characteristic time. As a result, the very flat particle spectra formed 
can extend to higher energies in comparison to the unperturbed 
conditions. 
 
One should also note that a realistic description of the acceleration 
process requires considering the full 3D configuration of the reconnection 
region (e.g. Birn \& Hesse 1994, Schopper et al. 1999, Veltri et al. 
1998). Besides the role which the mean field structure plays, particle 
transport depends in a qualitative way on the turbulence dimensionality 
(cf. Giacalone \& Jokipii 1994, Micha{\l}ek \& Ostrowski 1996) and the 
involved wave modes (e.g. Schlickeiser \& Miller 1997, Micha{\l}ek \& 
Ostrowski 1998). 
 
Besides providing a substantial improvement in understanding of the 
process, a general deficiency of the above mentioned papers discussing 
particle acceleration in the turbulent reconnection regions are the
complicated interrelations between such factors as the assumed or 
derived mean magnetic field structure, its dimensionality, the assumed 
or derived form of turbulence, and the particle acceleration process. 
For example in the paper of Ambrosiano et al. (1988) it was 
difficult to separate the role of diffusive particle motions from 
trapping by the structures formed in/near the reconnection layer. In 
order to clarify the situation, with only one of these factors modifying
the acceleration process in turbulent reconnection, we decided to  
use a simple model which allows us to consider the role of random 
particle trajectory perturbations separately. The process acting in the 
vicinity of the X-type reconnection does not provide a means for 
particle trapping in the mean (unperturbed) structure of the 
reconnecting field. As described in the next section, we use the 3D 
analytic model of stationary reconnection of Craig et al. (1995) to 
define the background model, being a reference for the model involving 
particle pitch-angle momentum scattering due to MHD turbulence. 
Perturbations of particle trajectories due to a turbulent magnetic field 
component were simulated using small-amplitude pitch-angle 
momentum scattering, enabling modelling of both small and large 
amplitude turbulence in a wide wave vector range. No second-order Fermi 
acceleration processes are allowed within this approach. In section 3 we 
present results of particle energy spectrum modelling for different 
scattering amplitudes. Comparison of the acceleration process to the 
unperturbed one confirms that the turbulence can substantially increase 
the acceleration efficiency, enabling particles to form flat high-energy 
spectra with much increased final energies. Then, in section 4, we 
briefly summarize these results. 
  
In the present simulations we scale the parameters of the reconnection 
region to those characteristic for the solar flare (e.g. Miller et 
al. 1997). However, we do not aspire to present a flare acceleration 
model. We provide all numerical values in SI units.

\section{Simulations of the acceleration process}  
  
\subsection{The 3D reconnecting magnetic field structure}  
  
\begin{figure}           
        \vspace{50mm}  
        \includegraphics{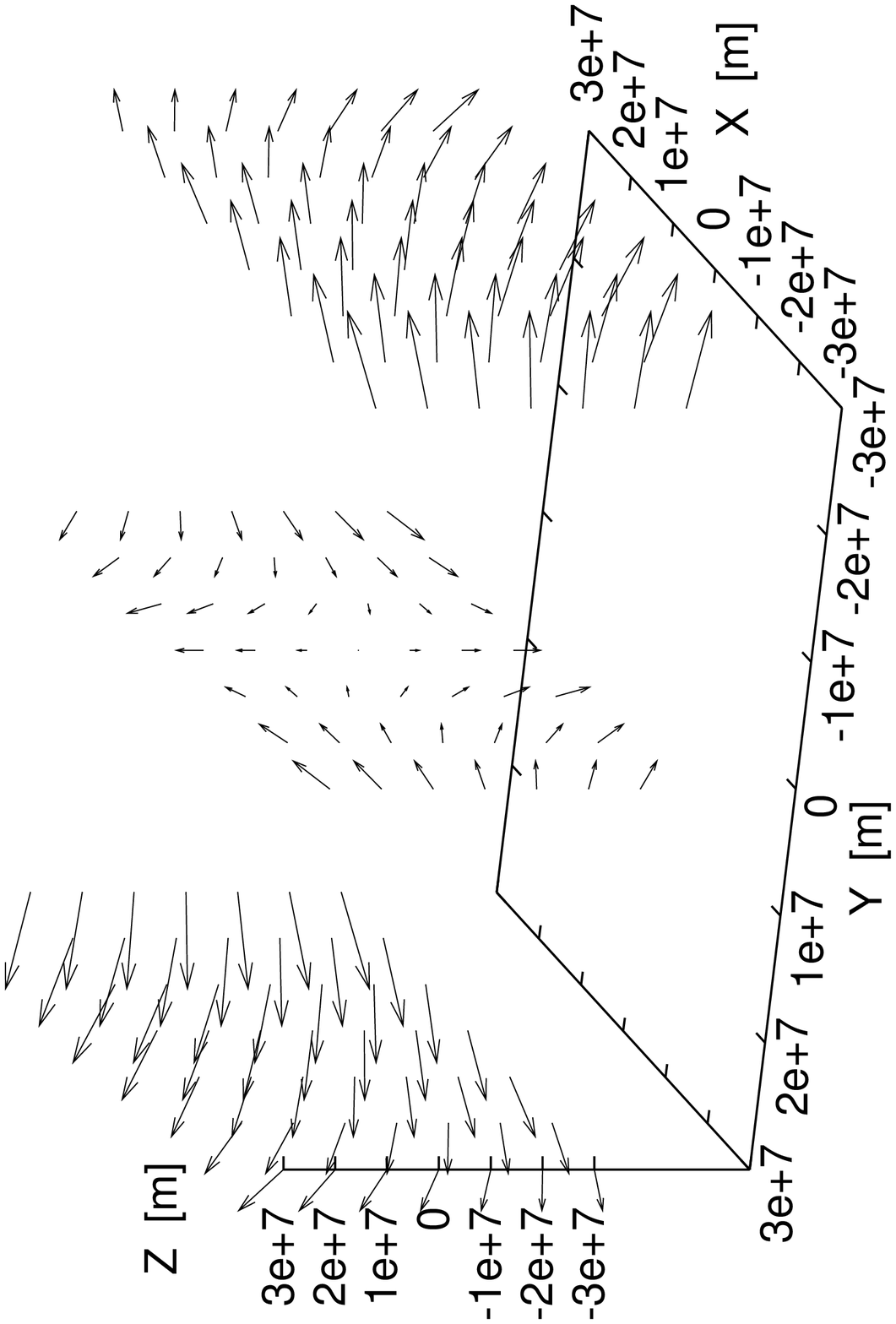}  
        \vspace{55mm}  
        \includegraphics{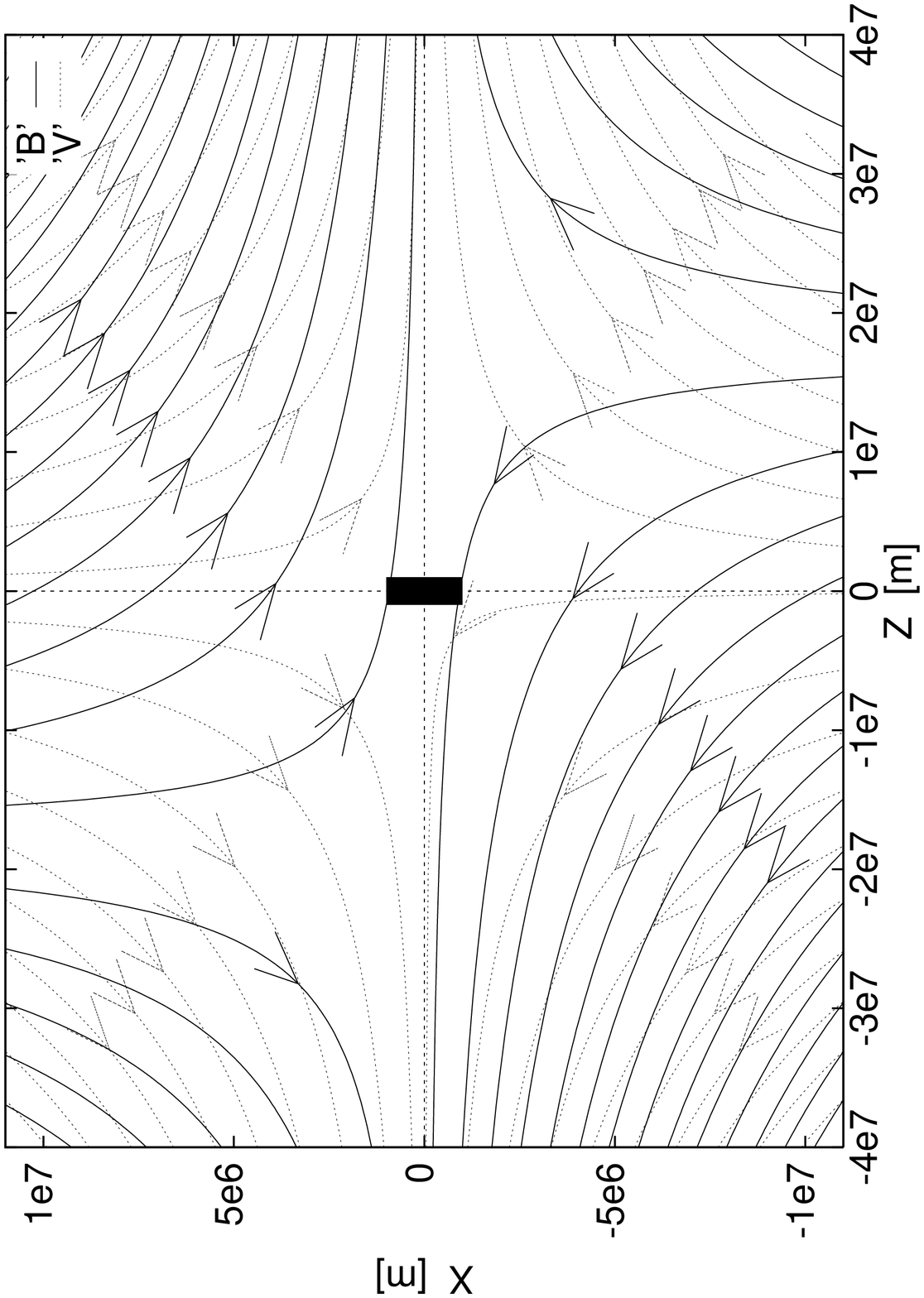}  
        \vspace{60mm}  
        \includegraphics{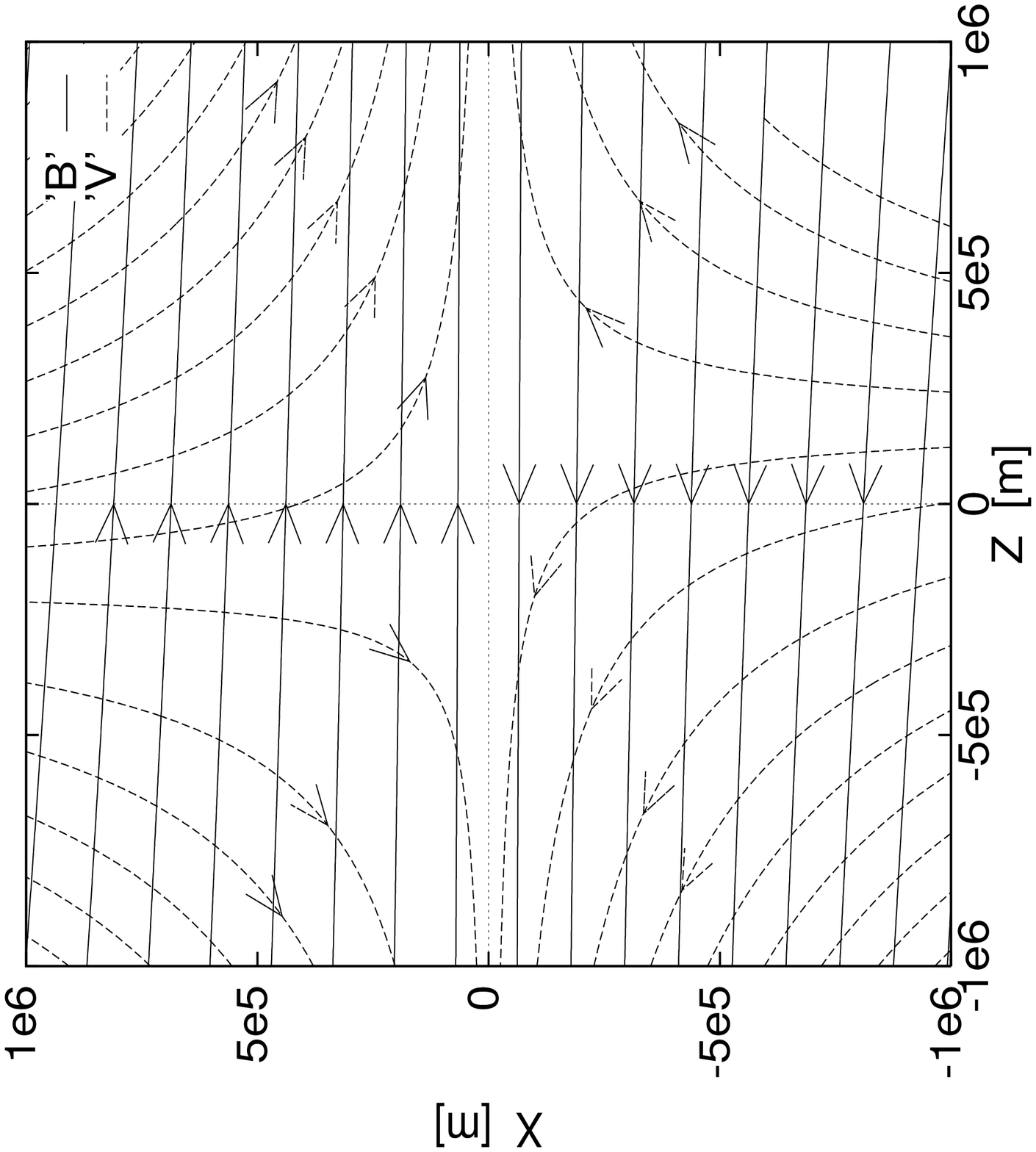}  
\vspace{13mm}
\caption{Upper panel: Magnetic field vectors in three layers 
$y=-3e7$, $y=0$, $y=3e7$ of the considered recconection volume. 
The y component of magnetic field vector has been 
increased 6 times to demonstrate of the stucture. 
Middle and 
bottom panels:
The magnetic field structure (full lines) and the velocity 
field (dashed lines) in the $y = 0$ plane 
of the considered reconnection region. In both these 
panels the border of the figure also represents the particle escape 
boundary. One should note different scales along the axes in these panels. 
The small black rectangle in the middle shows the size of the region 
where particle scattering due to turbulence is imposed. In the lower 
panel, the structure of the unperturbed magnetic field and the velocity 
field in this small turbulent volume is presented.} \end{figure} 
 
The considered reconnection region (Fig.~1) involves in-flowing plasma 
from the `top' and the `bottom' toward the $x = 0$ plane, and outflowing 
to the sides. All physical quantities, the flow velocity, the magnetic 
field and the electric field are symmetric with respect to the {\it 
origin} of the reference frame, with the reconnection ``plane'' 
(magnetic field is strictly 0 only in a single point in the center) 
being slightly inclined with respect to the $x = 0$ plane. To describe 
this structure analytically we use a 3D model of the stationary 
reconnection by Craig et al. (1995), which provides the analytic 
structure for the magnetic field $\vec{B}$, the electric field $\vec{E}$ 
and the plasma velocity $\vec{V}$. The following parameters are introduced 
in this model: $\alpha$, $\lambda$, $\eta$ and $k$  define 
topologic structure of the magnetic field; $\mu$ define the reconnection 
area thickness; $\rho$ is the plasma density; $\xi$  denotes the 
external electric field component; $\mu_0$ is the magnetic permeability 
of the vacuum. With the use of these parameters and the notation 
involving a radius vector $\vec{r} = \{ x$ , $y$ , $z \}$ and the unit 
vectors $\hat{x}$, $\hat{y}$ and $\hat{z}$ along the respective axes, 
  
$$ \vec{B} =  \lambda \vec{P}(\vec{r}) + \vec{Q}(x) \qquad , \eqno(2.1)$$  
  
$$ \vec{V} =  \left[ \vec{P}(\vec{r}) +  \lambda \vec{Q}(x) \right]{ 1  
\over \sqrt{\rho \mu_0}} \qquad , \eqno(2.2)$$  
  
$$ \vec{E} = \left[ -{\sqrt{\pi} \eta \over 2 \mu_0 } Z^\prime(0)  
\, e^{-(\mu x)^2} \right] \hat{y} \, +  
\xi \left[1 - 2 x \mu \, {\rm daw}(\mu x) \right] \hat{z} \qquad ,  
\eqno(2.3)$$  
  
\noindent  
where  
  
$$ \vec{P} = \alpha \left[ -x \hat{x} + k y \hat{y} + (1-k) z \hat{z}  
\right] \qquad , \eqno(2.4)$$  
  
$$ \vec{Q} =  {\xi \over \eta \mu}{ \sqrt{\mu_0 \over \rho}}  
 {\rm daw} ( \mu x ) \hat{y} + \left[ {  
\sqrt{\pi} \over 2\mu} Z^\prime(0) \, {\rm erf}(\mu x) + Z(0) \right]  
\hat{z} \qquad , \eqno(2.5)$$  
  
\noindent  
and  
$$ {\rm daw}(x) \equiv \int_0^x \exp (t^2 - x^2) dt \qquad , \quad 
   {\rm erf}(x) \equiv \int_0^x \exp (-t^2) dt \quad , $$ 
$$  \mu^2 = {\alpha \over 2 \eta}{1 - \lambda } \quad .   $$  
  
\noindent  
The function  $ Z(x) $ of Craig et al. (1995) satisfies the differential  
equation  
  
$$ \alpha (1-\lambda^2) [(1-k)Z + x Z^\prime] + \eta Z^{\prime \prime} =  
\alpha \gamma_{2} \lambda (2-k)x \quad , \eqno(2.6)$$  
  
\noindent  
where one requires $ \gamma_{2} k = 0 \quad $ ($\gamma_2$ or $k$ must be 
zero). With these expressions one can construct the reconnection region 
in two ways. It is possible to choose the parameters $\xi$ , 
$Z^\prime(0)$, $Z(0)$, $\alpha$, $\beta$, $\eta$, $\mu$ in such a way 
that either the magnetic field component $B_{\hat{z}}$ or $B_{\hat{y}}$ 
vanishes in some area: 
  
\begin{itemize}  
  
\item i.) $B_y$ vanishes where $\lambda \alpha k y = -{\xi \mu_0 \over  
\eta \mu } \, {\rm daw} (\mu x)$  
  
\item ii.) $B_z$ vanishes where  $\beta (1-k) z = {\sqrt{\pi} \over 2  
\mu} Z^\prime(0) \, {\rm erf} (\mu x)$.  
  
\end{itemize}  
  
\noindent  
The maximum value of $r$ allowing for one of the above conditions to be 
satisfied gives the half length of the reconnection region $L$ (a scale 
for the layer with negligible magnetic field near the null point). For 
example, for the case (ii) one has 
  
$$L = {\pi Z^\prime (0) \over 4 \beta (1-k) \mu } \qquad . \eqno(2.7)$$ 
  
\noindent
In the case (i), where the function $daw$ determines the behaviour of the
magnetic field, the gradient of the magnetic field is greater than in
the case (ii). As a result the simulation time for the required accuracy
is longer. Moreover, within the turbulent volume, at a large distance
from the centre the function $daw$ grows slower resulting in less
efficient acceleration. For these reasons in the present simulations we
consider solutions determined with the condition (ii) satisfied. We take
the following values of the model parameters: $\alpha = 1.0 \cdot
10^{-9}$, $\lambda = {\beta \over \alpha}$, where $\beta = 1.0 \cdot
10^{-10}$, $\eta = 1.0 \cdot 10^{-6}$, $k = 0.15$ , $\rho = M_p \cdot
10^{16}$ ($M_p$ proton mass), $\xi=0$, $Z(0)=0$,
$Z^\prime(0)=7 \cdot 10^{-4}$. Values of these parameters were 
derived from fitting the model magnetic, electric and velocity field to 
the mentioned solar flare conditions, with the plasma density equal to 
$10^{16}$ m$^{-3}$. 
 
Let us note that the simple analytic form of the considered solution 
limits the range of physically acceptable parameters to provide the 
possible boundary values of $\vec{V}$, $\vec{B}$, $\vec{E}$~. In the 
present simulations we choose such boundary values for the solution at 
the edge of our considered volume to provide physical parameters 
near the reconnection layer that are close to the ones estimated for the 
solar flares. In the vicinity of the reconnection area, just outside of 
the reconnecting current layer, the magnetic field induction is assumed 
to be $B = 1.68 \cdot 10^{-3}$~T and the respective gyroradius 
(gyroperiod) for a $1$ MeV proton is $r_g = 87$~m ($T_g = 4  \cdot 
10^{-5}$~s), and $r_g = 3.4$~km ($T_g = 8.1 \cdot 10^{-5}$~s) for a $1$ 
GeV proton. The assumed linear dimensions of the full considered 
reconnection region are $\Delta x = 2 \cdot 10^4$~km, $\Delta y= 2 \cdot 
10^5$~km and $\Delta z = 1.5 \cdot 10^5$~km. The size of the volume 
of the perturbed magnetic field is in the centre $\Delta x = 2 \cdot 
10^3$~km, $\Delta y = 6 \cdot 10^3$ km and $\Delta z = 2 \cdot 10^3$~km. 
One should note that the selected particular conditions are not 
essential for the presented considerations.

\subsection{Monte Carlo modelling procedure}  
  
In the simulations we used a Monte Carlo approach including a trajectory 
splitting technique to improve statistics at larger energies (cf. 
Ostrowski 1991). To derive particle spectra we recorded `weights' of 
particles escaping from the simulation volume in a given energy range. 
For every escaping particle a randomly\footnote{With probability of 
selecting a given particle being proportional to its weight.} chosen 
particle still active in the simulations was split into two identical 
particles, each with half the weight of the original particle. Then the 
trajectories of both particles were followed, but due to the applied 
random momentum scattering they evolve in different ways. The maximum 
simulation time $t_{max}$ ($t_{max}$ is the upper limit in Fig.~4) was 
chosen to be sufficiently large to be unreachable by high weight 
particles. Thus any further increasing of $t_{max}$ does not influence 
the resulting spectra in a visible way. 
  
In the simulations, we inject test particles with initial energy 
$E_0$ = $2$ MeV in the vicinity of the reconnection null point and we follow 
their trajectories by numerical integration of particle equations of 
motion. The integration is completed when a particle either crosses the 
boundary of the considered reconnection region (`escape boundary'), or 
the time limit $t_{max} = 100$~s is reached. Particles were scattered in 
a small region around the reconnection null point only. This region is 
shown in the two upper panels of Fig.~1 as black rectangles, and in 
expanded form in the lower panel. The trajectory computation times are 
much longer for low energy particles and this is the main reason why we 
start with initial proton energies substantially larger than the thermal 
energy; for the discussion of the injection problem of energetic solar 
flare particles one can consult the discussion in Miller et al. (1997). As 
the long integrations performed here require high accuracy we use a 
variable step fourth-order Runge-Kutta integration with the parameters 
chosen in such a way that any further accuracy increase does not affect 
the simulated trajectories. In the simulation we use only 100 particles 
because the required high accuracy of integration leads to extensive 
integration times. Let us remember that with the applied trajectory 
splitting technique the number of particles forming the spectrum at 
different energies is the same. 
  
We performed a number of numerical tests of the code applied in the 
simulations. We checked by hand the derivation of the resulting particle 
phase space co-ordinates in a few randomly chosen individual integration 
steps of the code algorithm. Then we derived particle trajectories 
within a few simple uniform magnetic fields oriented randomly with 
respect to the chosen reference frame. The results coincided within the 
numerical accuracy with the derived analytic trajectories. Finally, for 
the actual considered magnetic field structure in the reconnection 
volume we positively checked conservation of particle energy with the 
plasma velocity set equal to zero. 
  
In the simulations particles gain energy mostly in the vicinity of the 
reconnection layer when drifting in the $\vec{V} \wedge \vec{B}$ 
electric fields. Away from this region, while moving across the magnetic 
field gradient, particles can gain and lose energy, but the mean energy 
change is small. Somov \& Kosugi (1997) estimated the accelerated 
particle energy as a product of the magnetic field, the plasma velocity 
and the reconnection area length. The energy gains derived in our {\it 
unperturbed} reconnection model are in agreement with this estimate if 
we take the distance traversed by a particle within the reconnection 
area as the required length. The effect of particle escape from the 
reconnection volume has been discussed by Speiser  (1965) for a highly 
simplified reconnection model. We confirm his results showing that a 
small vertical magnetic field component within the reconnection layer 
increases the particle escape substantially. 
 
In the simulations we consider the particle scattering process only in a 
small volume containing the reconnection null point, where the magnetic 
energy is dissipated (see Fig.~1). The real reconnection regions are 
expected to show analogous structures, with the turbulence amplitude 
growing toward the centre, including null points of the magnetic field. 
This choice also resulted from the fact that the Craig et al. (1995) three 
dimensional model of the stationary reconnection is not too realistic 
at large distances from the null point, as, for example, it involves 
the inflow velocity growing without a limit when increasing distance 
from the centre. For the perturbed trajectories model we use a simple 
pitch-angle scattering approach (e.g. Ostrowski 1991) intended to model 
the particle scattering at MHD waves. In this case integration of the 
particle equations of motion is performed in the electromagnetic field 
defined by the unperturbed analytic model, but trajectory perturbations 
are introduced every constant time interval $\Delta t$, when the 
particle momentum vector is randomly scattered within a narrow cone 
along its original direction. The scattering is performed in the local 
{\it plasma} rest frame and it conserves particle energy in this frame. 
In the present simulations we consider the uniform momentum scattering 
within a cone with the half opening angle equal to $11^\circ$. The 
perturbation intensity is controlled by changing $\Delta t$ and it is 
characterized with $\aleph \equiv \kappa_\perp / \kappa_\|$, the ratio 
of the cross-field diffusion coefficient to the diffusion coefficient 
along the magnetic field. The value of $\aleph$ was determined in 
auxiliary simulations performed in the uniform magnetic field with the 
value characteristic for a region close outside the reconnection current 
sheet (see section 2.3). One should also note that decrease of the 
magnetic field toward the reconnection site leads to increasing the 
effective turbulence amplitude (i.e. the value of $\aleph$; in the limit 
of $B = 0$ we have $\aleph = 1.0$), but in the present simulations the
particle gyroradius is always larger than the reconnection layer 
thickness near the considered X-type null point. 
 
As the mean scattering time $\Delta t$ is assumed to be constant within 
a given simulation run, particle trajectories are affected by 
perturbations with intensity depending on particle energy and the 
background magnetic field. For non-relativistic particles with constant 
angular velocities of their gyration movements, assuming constant 
$\Delta t$ is equivalent to introducing scattering acts at constant 
gyrophase steps. Thus the resulting value of $\aleph$ does not depend on 
energy, as expected for the flat wave power spectrum $F(k) \propto 
k^{-1}$. This slightly unrealistic wave spectrum allows, on the other 
hand, evaluation of the role of diffusive effects for the same scattering 
amplitude at all considered particle energies. A discussion of a more 
realistic Kolmogorov wave spectrum within the finite wave vector range 
will be presented in the next paper (in preparation). However, as such
a wave spectrum carries more energy in long waves, the results are 
expected to show a transition from our low $\aleph$ results to the large 
$\aleph $ ones. 
 
In attempting to compare our simplified scattering model with the real 
turbulence with an amplitude of waves resonant for a particle 
of a given energy, $\delta B_r$ ($\delta B_r^2 / 8 \pi \approx F(k_r) 
\cdot k_r$, where $k_r = 2 \pi / r_g$), one can refer to a qualitative 
discussion of energetic particle diffusion presented by Drury (1983). 
With his scaling $\kappa_\| \propto (\delta B_r / B)^{-2}$ and 
$\kappa_\perp \propto (\delta B_r / B)^2$, the amplitude for resonance 
waves can be evaluated as $\delta B_r / B \approx \aleph^{1/4}$.

\subsection{Derivation of the $\aleph \equiv \kappa_\perp / \kappa_\|$  
parameter} 

The respective values of $\aleph \equiv \kappa_\perp / \kappa_\|$ were
derived in auxiliary simulations involving the spatially uniform
background magnetic field with the induction of $1.5 \cdot 10^{-3}$~T
and particles with energies equal to the initial energy $E_0 = 2$ MeV.
Trajectories of a large number of particles were followed with the
imposed scattering process involving the momentum angular scattering,
uniform within a cone of half opening angle equal to $11^\circ$ and with
the cone axis directed along the original momentum vector. The only
parameter varying between the simulations was the time interval between
successive scattering events, $\Delta t$. The resulting diffusion
coefficients were derived from growing particle dispersions along the
background field ($\kappa_\|$) and along two orthogonal axes
perpendicular to the background field ($\equiv$ along the $1$- or
$2$-axis). The results of such computations are presented in Fig.~2.
Presentation of two derived values of $\kappa_1 / \kappa_\|$ and
$\kappa_2 / \kappa_\|$ allows one to evaluate the accuracy of these
computations. The values used in the paper are fits to the
asymptotic $(T \to \infty)$  value of $\aleph = 0.5 \, 
(\kappa_1 + \kappa_2) / \kappa_\|$. For a sequence of scattering times 
$\Delta t$ = $10^{-6}$, $10^{-5}$, $10^{-4}$ and $10^{-3}$ 
we derived the respective 
values of $\aleph$ = $1.2 \cdot 10^{-2}$, $6 \cdot 10^{-3}$, $6 \cdot 
10^{-5}$, and $6 \cdot 10^{-7}$. The results derived without applying 
any scattering are indicated by $\aleph = 0$. 
  
\begin{figure}            
        \vspace{50mm}  
        \includegraphics{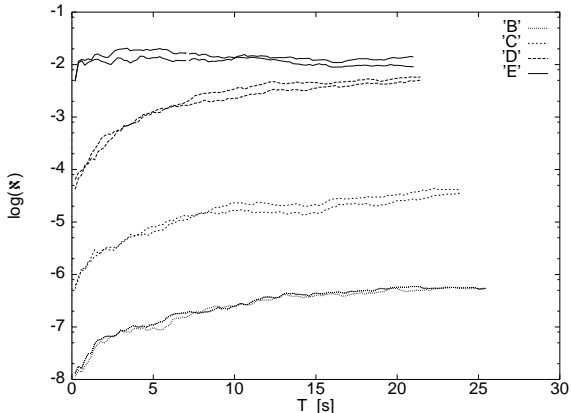}  
\vspace{9mm}  
\caption{The diffusion parameter $\aleph \equiv \kappa_\perp / \kappa_|$ 
calculated during particle distribution evolution. $T$ is the simulation 
time. For given initial parameters one has two resulting values of 
$\aleph$ for $\kappa_\perp = \kappa_1$ and $\kappa_\perp = \kappa_2$, 
presented as separate lines. } \end{figure} 
 
\begin{figure}             
        \vspace{50mm}  
        \includegraphics{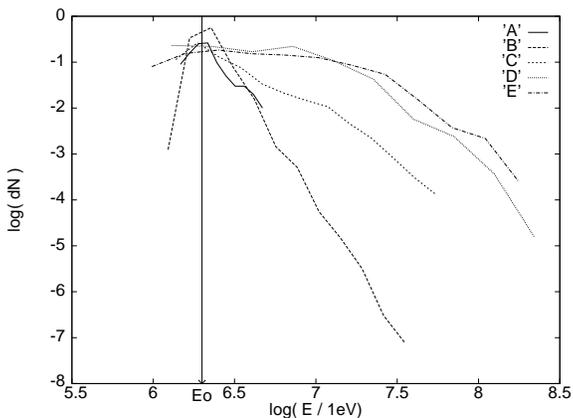}  
\vspace{9mm}  
\caption{The resulting energy spectra for protons injected at $E_0 = 2$ 
MeV: Model A: $\aleph = 0$ (the unperturbed model), Model B: $\aleph = 6 
\cdot 10^{-7}$, Model C: $\aleph = 6 \cdot 10^{-5}$, Model D: $\aleph = 
6 \cdot 10^{-3}$, Model E: $\aleph = 1.2 \cdot 10^{-2}$. A vertical $E = 
E_0$ line is provided for a reference.} \end{figure}

\section{Spectra of accelerated particles}  
  
In order to provide qualitative evaluations of turbulence effects in the 
volume of reconnecting magnetic field, but considering it only as a 
factor introducing random motion component to particle trajectories, we 
performed simulations of energetic proton spectra with a varying amount of 
turbulence ($\equiv$ scattering). As explained above, this approach 
assumes the existence of short wave magnetic field perturbations to be 
present in the limited volume -- the black rectangle in Fig.~1 -- 
near the central neutral point, but we do not consider the influence of
the turbulence on the reconnection process. Thus it is a complementary 
approach to that using MHD modelling of the turbulent 
reconnection including wave perturbations from a narrow wave vector 
range, as discussed in section 1. 
 
We performed simulations of particle evolution starting at the same 
`injection' energy $E_0 = 2$ MeV, avoiding consideration of the real 
injection process at much lower energies (cf. Miller et al. 1997). For 
each set of particles we derived the spectrum of particles escaping from 
the reconnection volume, as illustrated in Fig.~3. In the non-perturbed 
model ($\aleph = 0$, curve A) protons can increase their initial energy 
by approximately $70$\%~. One should note that the injected energetic 
particles can gain as well as loose energy. Introducing 
trajectory perturbations results in substantial modification of the 
acceleration process (curves B, C, D, E in Fig.~3). The spectrum energy 
cut-off shifts to higher values and the spectrum becomes harder when the 
amount of scattering (`turbulence amplitude') is increased. In our 
simulations the resulting flat spectra extend up to $50$ MeV for models 
with strong turbulence (Model D and E), and a steeper part of the 
spectrum is recorded at energies above $100$ MeV. This behaviour results 
from the fact that the diffusive component introduced in to particle 
trajectories by the scattering enables some particles to stay in the 
reconnection region much longer and diffuse back close to the null 
point from outside. As illustrated, this can have a pronounced influence 
on the acceleration process by substantially increasing the particle mean 
energy gain and providing much larger energies of individual particles. 
There is a general trend for the acceleration efficiency to increase 
with the perturbation amplitude in the considered range of values for 
the $\aleph$ parameter.

\begin{figure}           
        \vspace{50mm}  
        \includegraphics{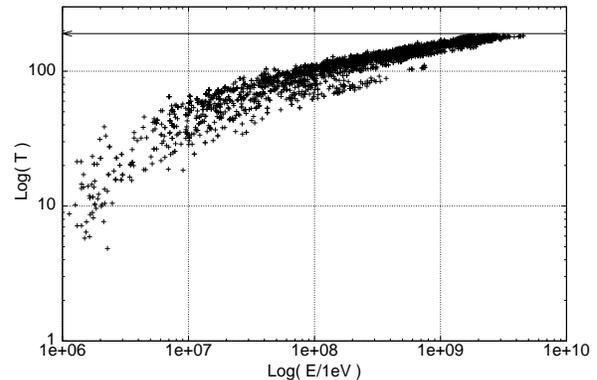}  
\vspace{9mm}  
\caption{Acceleration times, $T$, versus the final particle energies, 
$E$, in the Model C (cf. Fig.~3). The simulation time cut-off at 
$t_{max} = 100$~s is indicated with a dashed line. Due to the applied 
trajectory splitting procedure the particles represented by points close 
to $T = t_{max}$ have very small weights and further increasing 
$t_{max}$ does not lead to modification of the spectra presented in 
Fig.~3.} \end{figure} 
  
Acceleration by the uniform electric field is characterized with the 
energy-independent rate of particle energy increase, leading to a 
linear relation between the particle acceleration time and its final 
energy. In contrast, a plot of correlation between the individual 
particle acceleration time ($\equiv$ the physical simulation time) and 
its final energy presented in Fig.~4 shows that the acceleration rate 
depends on energy, being preferentially defined by particle drifts in 
the `$V$ {\rm x} $B$' electric fields. A large energy dispersion in a 
given time reflects variety of involved particle diffusive trajectories. 
 
\section{Final remarks} 
 
Consideration of energetic particle acceleration accompanying the 
magnetic field reconnection process requires knowledge of the 
electromagnetic field structure in the region of interest. Until now the 
available numerical models of such fields were oversimplified and fail to 
consistently include together the short and long MHD wave modes. 
Therefore, in the present simulations we apply an analytic model for a 
reconnecting field as the background, and the particle scattering 
process is imposed as the small amplitude uncorrelated angular momentum 
perturbations. A comparison of the acceleration process in such a model 
to the unperturbed model reveals the important result that inclusion of 
the turbulent field component into the reconnection volume can change 
the acceleration process in a {\it qualitative} way, enabling particles 
to reach much higher final energies and significantly increase their 
mean energy gain. 
  
A serious limitation on the validity of our simplified modelling arises 
due to a non-self-consistent introduction of the MHD turbulent field. 
The trajectory perturbations in the turbulent medium are considered 
without taking into account the influence -- in both directions -- of 
these turbulent motions on the reconnection process itself. As mentioned 
in section 1, the presence of magnetic field turbulent structures 
increases mixing in the medium and is the source for anomalous 
resistivity leading to more effective reconnection as discussed 
recently by Lazarian \& Vishniac 2000. Subsequently, it 
leads to more efficient particle acceleration (Matthaeus et al. 1984, 
Ambrosiano et al. 1988, Scholer \& Jamitzky 1989, Veltri at al.  1998, 
Birn \& Hesse 1994, Kliem et al. 1998, Schopper et al. 1999). As 
discussed in the present paper from a slightly different perspective it 
is expected to increase the cosmic ray acceleration efficiency, and 
influence the involved time scales and details of the resulting energy 
spectrum. However, for a given scattering conditions within the 
reconnection region the spectrum upper energy cut-off is limited by the 
`global' perturbed structure of the reconnecting volume and not by the 
local conditions within the thin reconnection current sheet. 
  
The present work was supported by the {\it Komitet Bada\'n Naukowych} 
through the grants PB 179/P03/96/11 and PB 258/P03/99/17.

\end{document}